\documentclass[prl,twocolumn,amsmath]{revtex4}

\usepackage{graphicx}
\usepackage{amssymb,amsfonts,amsmath}

\def \>{\rangle} 
\def \<{\langle} 
 
\def\be{\begin{equation}} 
\def\ee{\end{equation}} 
\def\longrightharpoonup{\relbar\joinrel\rightharpoonup}
\def\longleftharpoondown{\leftharpoondown\joinrel\relbar}

\def\longrightleftharpoons{
  \mathop{
    \vcenter{
      \hbox{
      \ooalign{
        \raise1pt\hbox{$\longrightharpoonup\joinrel$}\crcr
	  \lower1pt\hbox{$\longleftharpoondown\joinrel$}
	  }
      }
    }
  }
}

\newcommand \bea {\begin{eqnarray}} 
\newcommand \eea {\end{eqnarray}}

\DeclareMathOperator{\Tr}{Tr}

\begin{document}

\title{Comment on ``Why does deep and cheap learning work so well?"}

\author{David J. Schwab}
\email{david.schwab@northwestern.edu}
\affiliation{Department of Physics and Astronomy,
  Northwestern University, Evanston, IL 08854} 
\author{Pankaj Mehta}
\email{pankajm@bu.edu}
\affiliation{Department of Physics, Boston University, Boston, MA
  02215}

\begin{abstract}
In a recent paper, ``Why does deep and cheap learning work so well?", Lin and Tegmark claim to show that the mapping between deep belief networks and the variational renormalization group derived in   \cite{mehta2014} is invalid, and present a ``counterexample'' that claims to show that this mapping does not hold. In this comment, we show that these claims are incorrect and stem from a misunderstanding of the variational RG procedure proposed by Kadanoff. We also explain why the ``counterexample'' of Lin and Tegmark is compatible with the mapping proposed in  \cite{mehta2014}.
\end{abstract}
\maketitle

In this brief comment, we respond to claims in \cite{lin2016} suggesting that the mapping derived between the variational RG procedure and deep belief networks is invalid. 
We will not comment on the other claims about the relationship between RG and neural networks made in the paper.

The central criticism in \cite{lin2016} to our work is that a coarse-graining transformation that preserves the free energy does not suffice to reconstruct the empirical (microscopic) probability distribution. In an appendix, they provide a concrete example of this. We entirely agree with this statement and \emph{never claimed otherwise}. 
 
In fact, this is not a new observation and was already well-known to Kadanoff and collaborators working on variational RG. For example in \cite{kadanoff1976}, they state:
\begin{quote}
Hopefully, one might obtain good results for physical quantities by choosing the upper (lower) bound recursions that give the smallest error in the free energy... We say "hopefully" because usually one is not interested in the free energy itself. Rather its derivatives are of the major physical interest. Since the variational principles pertain to the free energy, there is no guarantee that the derivatives will be accurate. 
\end{quote}
The underlying reason for this is straightforward. In RG, one often aims to preserve the long wavelength part of a distribution. Preservation of the free energy, a single number, is entirely insufficient to yield useful results. Since physical quantities usually depend on the derivatives of the free energy (i.e. magnetization in the Ising model is the derivative with respect to the field), any reasonable RG procedure must preserve more information about the distribution than just the free energy.

In variational RG, this is achieved by the replacing the preservation of the free energy with a ``trace condition'' on the renormalization operator, $T(\bf{v},\bf{h})$, relating the original (visible) degree of freedom $\bf{v}$ and the coarse grained (hidden) degrees of freedom $\bf{h}$:
\be
\Tr_{\bf h} e^{T(\bf{v},\bf{h})}=1
\label{eq1}
\ee
The trace condition above implies that the free energy is preserved. In fact, as discussed in   \cite{mehta2014}, the trace condition is so strong that it implies that the microscopic probability distribution is unperturbed by the introduction of the auxiliary variables. 

However, the converse is not necessarily true. It is possible to preserve the free energy while violating the trace condition (note the typo in eq. (8) of \cite{mehta2014} -- it is not a biconditional. This typo possibly contributed to this misunderstanding). The example in appendix A of \cite{lin2016} is precisely an example of a transformation that violates the trace condition but preserves the free energy (see below). For this reason, the example of \cite{lin2016} does not invalidate the mapping between variational RG and deep belief networks derived in  \cite{mehta2014}.

 We hope this brief comment clarifies the root of the misunderstanding in  \cite{lin2016}.  
 \appendix
 \section{Appendix: Trace condition is violated by ``counterexample'' of Lin and Tegmark}
In their appendix, Lin and Tegmark construct a model that consists of visible (microscopic) degrees of freedom $\bf y$ and hidden degrees of freedom $\bf y'$. The original distribution of visible variables is $p({\bf y})=\frac{1}{Z}e^{-H({\bf y})}$, with $Z = \sum_{\bf y}e^{-H({\bf y})}$. The joint distribution is given by $p({\bf y},{\bf y'}) = \frac{1}{Z_{\text{tot}}}e^{-H({\bf y},{\bf y'})}$, where $H({\bf y},{\bf y'}) = H({\bf y}) + H({\bf y'}) + K({\bf y}) + \ln \tilde{Z}$, $Z_{\text{tot}}=\sum_{\bf y, y'}e^{-H({\bf y},{\bf y'})}$, and $\tilde{Z} = \sum_{\bf y}e^{-H({\bf y})-K({\bf y})}$. Thus, the coupling operator $-T({\bf y},{\bf y'}) = H({\bf y'}) + K({\bf y}) + \ln \tilde{Z}$. Importantly, they require $K({\bf y})$ to be non-constant. This immediately implies that $\Tr_{\bf y'} e^{T(\bf{y},\bf{y'})}$ is non-constant, and thus the trace condition in Eq. \ref{eq1} is not satisfied for any of the distributions they consider. 
 
\bibliography{refsmain}   
\end{document}